# Self-referenced, drift-tolerant dipole-resolved population inversion using degeneracy-lifted dual quasinormal modes

JIAXIN YU(于佳鑫)[1,**], XINYU ZHANG(张馨雨)[1], GUANGYU DAI(戴光宇)[1], SHUAI XING(邢帅)[1], MINGHUI YANG(阳明辉)[1], AND FUXING GU(谷付星)[1,**]

[1]*Laboratory of Integrated Opto-Mechanics and Electronics, School of Optical-Electrical and Computer Engineering, University of Shanghai for Science and Technology, Shanghai 200093, China*

[**]*Corresponding authors. Email: yujiaxin@usst.edu.cn, gufuxing@usst.edu.cn*

**Abstract:** Photoluminescence intensity is widely used to infer exciton populations, yet the detected signal inherently convolves occupancy with radiative-rate modification and collection efficiency, making quantitative inversion vulnerable to pump and system drifts. Here we realize a dual-channel self-referenced scheme enabled by two nearly degenerate quasinormal modes in a hybrid microcavity. Their shared optical path provides common-mode observables (i.e., overall spectral and intensity drift) that track global thermo-optic and pump fluctuations, while their differential-mode observables (i.e., spectral splitting and mode-contrasted emission) remain highly sensitive to local gap dielectric perturbations and dipole-dependent radiative weights. Using temperature as a control parameter in monolayer $WSe_2$, we exploit this common/differential-mode framework to robustly invert the relative populations of excitons with out-of-plane ($\perp$) and in-plane ($\parallel$) dipole transitions without external absolute calibration. At the temperature of ~50 K, we obtain $N_\perp/N_\parallel \approx 200$, coincident with the expected accumulation in the out-of-plane–emitting dark manifold. This internally referenced approach provides a practical route to drift-tolerant, dipole-resolved population metrology in nanogap photonic systems.

## 1. Introduction

The quantitative readout of exciton populations and distributions underpins the identification of exciton thermalization and non-equilibrium steady states, and further affects key physical processes such as coherence buildup [1] and condensation thresholds [2–4]. However, in most experiments, exciton populations are indirectly inferred from photoluminescence (PL) intensity. Under steady-state conditions, the detected signal is generally expressed as $I_{det} \propto N \times \Gamma_{rad} \times \eta_{col}$, where $N$ denotes the exciton populations, $\Gamma_{rad}$ the radiative rate, and $\eta_{col}$ the collection efficiency. Any variations in $\Gamma_{rad}$ or $\eta_{col}$—arising from mode coupling, radiation directionality, collection numerical aperture (NA), or alignment drifts—may be misinterpreted as changes in the exciton populations. In addition, pump fluctuations and optical-path misalignment introduce common-mode intensity drift, whereas fluctuations in the dielectric environment such as temperature drift and local geometric variations may simultaneously modify resonance conditions and radiative channels, thereby further undermining the reproducibility and traceability of intensity-based inversions.

This challenge is particularly severe for out-of-plane dipoles such as spin-forbidden dark excitons, whose emission predominantly occupies high–in-plane-wave-vector (high–$k_\parallel$) channels outside the detection NA, resulting in vanishingly low far-field radiation efficiency [5]. In recent years, manipulating the local density of optical states (LDOS) via micro/nanocavities [6–8] or near-field probes [9–11] has enabled the redirection of high-$k_\parallel$ emission to the far field for detection. However, reliable quantification of the exciton populations remains a bottleneck: The absolute PL intensity of a single enhancement mode often superimposes contributions from various exciton states and their respective coupling pathways, making it difficult to uniquely attribute intensity variations to changes in the exciton populations alone [12,13]. In addition, drifts in pump power, mechanical stability, and collection efficiency introduce systematic errors during long-term or dynamic measurements [14,15]. These considerations motivate the development of a self-referenced detection scheme integrated within the same optical platform—one that shares common-mode perturbations yet provides a differential response via distinct radiation selectivity, thereby enabling the decoupling of instrumental drifts from intrinsic population evolution.

In our previous work, we demonstrated that a system hybridized by surface plasmon polariton (SPP) and whispering-gallery mode (WGM) serves as an efficient far-field interface [16], which is composed of a $SiO_2$ microsphere (MS) on top of Au substrates (MS/Au, Fig. 1(a)). A dominant pair of near-degenerate quasinormal-modes (QNMs) can fold and redirect the high-$k_\parallel$ emission of out-of-plane dipoles on the substrate surface into collectable angular domains for quantitative detection. Building upon this, we further find that this structure naturally emerging from the degeneracy lifting of the mode pair provides a dual-channel physical foundation for self-referenced metrology. Specifically, the common-mode quantities, including the collective intensity and global spectral shift of both modes, track pump fluctuations and slow system drifts. The differential-mode quantities encode specific information: one quantity is the spectral splitting of the two modes, which is highly sensitive to variations in the local effective dielectric environment; and the second one is their discrepancy in selective enhancement of orientated dipole transitions, which enables the inversion of the out-of-plane (OP, represented by $\perp$) to in-plane (IP, represented by $\parallel$) population ratio. This dual-channel self-referenced readout provides a robust pathway for the quantitative characterization of weakly radiative states under dynamic conditions.

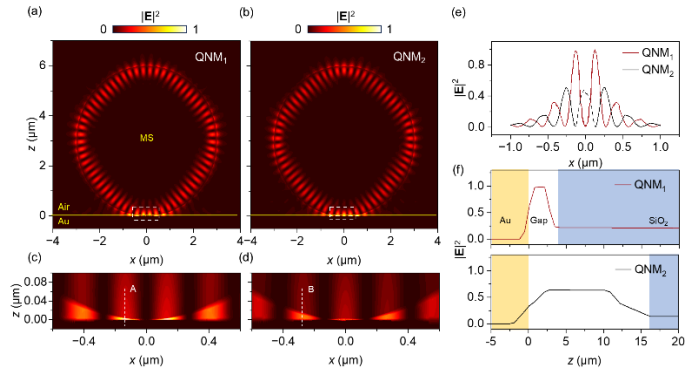

**Fig. 1.** Analysis on the electric-field distributions $|\mathbf{E}|^2$ of the two modes. (a,b) Field distribution of $QNM_1$ (a) and $QNM_2$ (b) in the $x$–$z$ cross-sectional plane, where $x$ and $z$ denote the lateral and vertical coordinates. (c,d) Enlarged views of the gap region for $QNM_1$ (c) and $QNM_2$ (d). A in (c) and B in (d) mark the positions of maximum $|\mathbf{E}|^2$ within gap for $QNM_1$ and $QNM_2$, respectively. (e) Line cuts along $z = 0$ for $QNM_1$ (red) and $QNM_2$ (black). (f) Line cuts along the $z$ direction at positions A for $QNM_1$ (top) and at positions B for $QNM_2$ (bottom). Shaded regions denote Au (yellow), gap layer (white), and $SiO_2$ microsphere (blue).

## 2. Results and Discussion

Simulation is performed using a full-wave finite-element calculations within a Green-tensor QNM framework [17–19], and device fabrication follows the microsphere-assisted $WSe_2$/Au configuration described in Ref. 16. The optical response of the WGM–SPP hybrid microcavity is governed by two nearly degenerate TM-like QNMs, denoted as $QNM_1$ and $QNM_2$. Although they share the same radial order and similar resonance frequencies, their field participation in the air gap between the MS and the substrate is markedly different (Fig. 1(a) and 1(b)). In first-order perturbation theory [20,21], the resonance shift scales with the field-energy weight in the perturbed region; therefore, the spatial distributions of $|\mathbf{E}|^2$ determine the relative dielectric sensitivities of the two modes. Because the field intensity of both QNMs is predominantly confined to the MS–substrate air gap, their sensitivity is governed by the gap field distribution (Fig. 1(c) and 1(d)). For example, at the immediate substrate surface, the $|\mathbf{E}|^2$ intensity of $QNM_1$ is nearly twice that of $QNM_2$ (Fig. 1(e)). Moreover, the first $|\mathbf{E}|^2$ maximum of $QNM_1$ is located in the lateral non-contact gap regions (marked by A in Fig. 1(c)), whereas that of $QNM_2$ concentrates at the contact center ($x = 0$, dashed curve in Fig. 1(e)). In addition, field cross-sections extracted at positions A and B (the second maximum of $QNM_2$, Fig. 1(f)) reveal that the electric field of $QNM_1$ varies more sharply within the gap, whereas $QNM_2$ changes more gradually. Consequently, $QNM_1$ has a substantially larger participation factor and is therefore highly sensitive to local perturbations, while the weakly perturbed $QNM_2$ can serve as a reference mode.

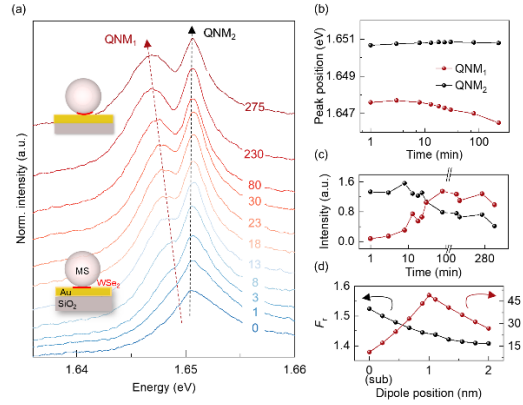

**Fig. 2.** Lifting the mode degeneracy via optical heating. (a) Time-resolved emission spectra. Curves are vertically offset for clarity. Dashed lines mark the resonance energies of $QNM_1$ (red) and $QNM_2$ (black). Numbers indicate acquisition time (min). Insets: schematics of the evolving bending of the active layer within the gap. (b,c) Peak positions (b) and intensities (c) of $QNM_1$ (red) and $QNM_2$ (black) versus time. (d) Simulated radiative-rate modification factor $F_r$ versus emitter coordinate $z$ for $QNM_1$ (red) and $QNM_2$ (black).

Experimentally, to locally modify the gap dielectric environment, we insert a monolayer $WSe_2$ into the gap and induce gradual bending of the monolayer via optical heating. Importantly, the optically induced bending predominantly modifies the non-contact gap regions, whereas the contact center is pinned (Fig. 2(a) inset); this makes the perturbation overlap much larger for $QNM_1$ than for $QNM_2$. As the deformation proceeds, the initially degenerate resonance evolves into two spectrally resolved peaks (Fig. 2(a)). Double-Lorentzian fitting indicates that the high-energy branch remains nearly unchanged during modulation, indicating that it corresponds to the weakly perturbed $QNM_2$, whereas the low-energy branch undergoes a cumulative redshift of approximately 1.3 meV, consistent with the expected higher dielectric participation of $QNM_1$ (Fig. 2(b)). The concurrent but distinct intensity evolution of the two peaks further supports the mode assignment. For a given pump and detection condition, the measured peak intensities $I(t)$ roughly reflect the cavity-modified radiative channels (Fig. 2(c)); we therefore compare their trends with the calculated, radiative enhancement quantified by the Purcell factors $F_r$. During deformation, the emission intensity of the high-energy mode decreases monotonically (Fig. 2(c), black), consistent with the calculated evolution of the radiative enhancement into $QNM_2$ for the corresponding dipole channel (Fig. 2(d), black, discussed later). In contrast, the low-energy mode exhibits a non-monotonic evolution— initially increasing and subsequently decreasing (Fig. 2(c), red)—matching the calculated trend for QNM$_1$ (Fig. 2(d), red). Combining the distinct dielectric sensitivity and the simulated intensity trends, we assign the low- and high-energy branches to QNM$_1$ and QNM$_2$, respectively, lifting the near-degeneracy of the two modes in a controlled manner. The bending-induced symmetry breaking therefore converts an otherwise hidden local perturbation into a measurable spectral splitting, establishing QNM$_2$ as a built-in reference and QNM$_1$ as the sensitive readout channel [22–24].

Beyond their frequency response, QNM$_1$ and QNM$_2$ display distinct polarization characteristics and far-field angular distributions. Polarization-resolved spectra indicate that QNM$_2$ retains a stable TM-like polarization (Fig. 3(a), black) and forms a narrow radiation band centered around ± 30° in energy-momentum ($E$–$k$) maps, with a full width at half maximum of ~ 5° (Fig. 3(c)), consistent with numerical simulations (Fig. 3(b), bottom panel). In contrast, QNM$_1$ is nearly unpolarized (Fig. 3(a), red), with radiation primarily confined within ± 20° near the optical axis (Fig. 3(d)), in agreement with simulation (Fig. 3(b), top panel). The inset of Fig. 3(c) and Fig. 3(d) further show that QNM$_2$ resonance energy is nearly dispersionless within the collected angular range, whereas QNM$_1$ exhibits a broadened, approximately parabolic dispersion. The dispersion can be understood by the strong dielectric participation of QNM$_1$ in the non-contact gap region: the spatially inhomogeneous dielectric perturbation predominantly affects QNM$_1$, leading to polarization mixing and dispersion broadening; whereas QNM$_2$ remains weakly perturbed, preserving its TM-like character and off-axis emission.

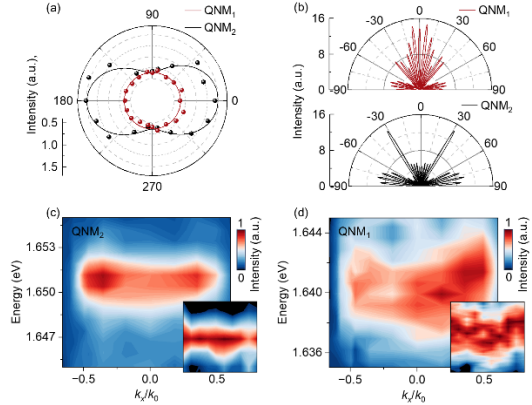

**Fig. 3.** Polarization- and $k$-resolved signatures of the degenerate-lifted modes. (a) Polarization-dependent intensity of QNM$_1$ (red) and QNM$_2$ (black). (b) Simulated far-field angular radiation patterns of QNM$_1$ (top) and QNM$_2$ (bottom). (c,d) Energy–momentum maps for QNM$_2$ (c) and QNM$_1$ (d), where $k_x$ denotes the in-plane wave-vector component and $k_0 = \omega/c$. Insets: each energy spectrum is normalized to its own maximum at the corresponding $k_x/k_0$.

This asymmetry naturally enables a self-referenced two-channel readout, which we first describe using differential-mode observables. For example, the frequency difference, $\Delta\omega = \omega_1 - \omega_2$, is highly sensitive to local symmetry breaking and effective permittivity variations within the gap (Fig. 2). To examine the pronounced sensitivity of QNM$_1$ to changes in the local dielectric environment, emission spectra are also extracted from different regions exhibiting varying bending amplitudes. Distinct differences in the frequency splitting between QNM$_1$ and QNM$_2$ are observed (Fig. S1). Meanwhile, the common-mode observables averaged frequency $\bar{\omega} = (\omega_1 + \omega_2)/2$ tracks global drifts (Fig. 4), whereas the ratio of the mode intensity $R = I_1/I_2$ rejects pump-induced common-mode fluctuations. To validate the self-referenced nature, we modulate the pump power, but the QNM$_1$/QNM$_2$ intensity ratio remains constant within uncertainty (Fig. S2), thereby confirming rejection of pump-induced common-mode fluctuations.

The established dual-channel self-referenced scheme combined with the distinct orientation-dependent radiative weights of the two modes allow us to extract the relative populations of oriented dipoles, i.e., $N_i$ ($i = \perp, \parallel$). The orientation-dependent coupling is quantified by the radiative enhancement factor $F_{r,i} = \Gamma_i/\Gamma_0$ [25], where $\Gamma$ and $\Gamma_0$ denote the radiative decay rates in the cavity and free space, respectively. $F_{r,i}$ is obtained from the same QNM-based response analysis as in Fig. 2(d). Specifically, QNM$_1$ exhibits strong selectivity toward OP dipoles, with $F_{r,\perp} : F_{r,\parallel} \approx 7.2$, whereas QNM$_2$ enhances both orientations comparably, with $F_{r,\perp} : F_{r,\parallel} \approx 1.98$.

To experimentally confirm the different orientation-dependent weight of the two modes, we tune the relative dipole populations via varying the global temperature. In monolayer WSe$_2$, the energy level of spin-forbidden dark excitons (primarily OP oriented) typically lies ~40 meV below that of bright excitons (primarily IP oriented), leading to an increasing ratio of $N_\perp/N_\parallel$ upon cooling [26–28]. Figure 4(a) presents the emission spectra from 297 K down to 7 K. Both QNMs exhibit a slight synchronous blueshift ($\bar{\omega}$ changes, Fig. 4(b)), while the splitting $\Delta\omega$ remains nearly unchanged, consistent with a common-mode thermo-optic drift rather than a differential-mode geometric deformation. As the temperature decreases, the direct emission (without antenna effect of the MS, dashed curve marked with $X_B$) bright excitons weakens upon cooling, yet the direct emission of dark excitons remains limited (dashed curve marked with $X_D$) due to the restricted NA of the detection system. However, near the dark-state line (~1.69 eV), the emission intensity of QNM$_1$ increases more rapidly than that of QNM$_2$, especially at the temperature lower than 50 K (Fig. 4(c)). In contrast, the two modes near the local-state line (~1.65 eV) show comparable intensity. The observation confirms the different orientation-dependent enhancement of the two modes.

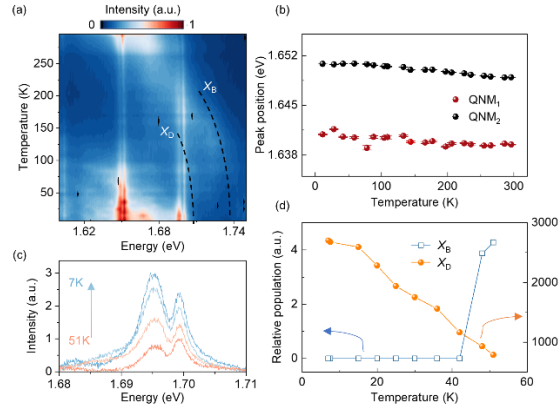

**Fig. 4.** Quantitative retrieval of dipole populations. (a) Temperature-resolved emission intensity map. The dashed guide lines trace the peak positions of bright ($X_B$) and dark excitons ($X_D$). (b) Peak positions of QNM$_1$ (red) and QNM$_2$ (black) versus temperature. (c) Representative emission spectra taken in the temperature range of 51 K to 7 K. The left, broad peak centered at ~1.695 eV is QNM$_1$ and the right, narrow peak centered at ~1.70 eV is QNM$_2$. (d) Relative populations of $X_B$ (blue, open symbols) and $X_D$ (orange, solid symbols) versus temperature.

In the low-temperature regime (i.e., 51–7 K), the energies drift of both bright and dark excitons became negligible [29], so the exciton-QNM detuning is taken constant. It allows the quantitative retrieval of the real-time dipole populations as the temperature varies. A quantitative model is then constructed as the sum of intensity contributions from the two dipoles orientation $I_{\text{det}}(T) \propto \sum_{i \in \perp, \parallel} N_i(T) \times \Gamma_{\text{rad},i} \times \eta_{\text{col}}$, where $N_\perp(T)$ and $N_\parallel(T)$ denote the populations of out-of-plane and in-plane excitons at temperature $T$, respectively. The cavity-modified radiative decay rate is given by $\Gamma_{\text{rad},i} = F_r \Gamma_{0,i}$, where $F_r$ is approximated by its maximum value determined by a full-wave simulation [16]. In monolayer $WSe_2$, the intrinsic radiative decay rates satisfy $\Gamma_{0,\perp} \approx 10^{-2} \Gamma_{0,\parallel}$ [26]. The collection efficiency $\eta_{\text{col}}$ represents the far-field collection coefficient of the mode in our self-referenced analysis, which is temperature-independent. Fitting of $I_{\text{det}}(T)$ shows that the OP population increases continuously upon cooling, whereas the IP population gradually decreases and becomes nearly negligible below 40 K (Fig. 4(d)), in overall agreement with previous reports [29,30]. This behavior reflects the relaxation and accumulation of excitons into the lower-energy dark manifold at low temperature. Notably, at T ≈ 50 K, we obtain an experimentally extracted ratio $N_\perp/N_\parallel \approx 200$. If one assumes a Boltzmann distribution, this ratio corresponds to an apparent energy scale of ~22 meV, substantially smaller than the commonly quoted ~40 meV splitting. This discrepancy points to a non-thermal steady state, suggesting incomplete thermalization between the two manifolds [31], likely due to suppressed inter-manifold scattering in this temperature range.

## 3. Conclusion

In summary, we demonstrate a dual-channel, self-referenced scheme enabled by two near-degenerate QNMs in a WGM–SPP hybrid microcavity, where common-mode observables track global spectral/intensity drift while differential-mode observables encode local gap dielectric perturbations and dipole-dependent radiative weights. Compared with single-resonance enhancement approaches that rely on absolute intensity and are therefore susceptible to pump/collection drifts and multi-path contributions, our architecture provides an internal reference channel that stabilizes intensity-based inversion. In contrast to scanning near-field methods, it remains compatible with long-duration and dynamic measurements because the reference and readout channels are acquired simultaneously through the same optical path. Looking forward, this self-referenced dual-mode metrology can be generalized to other nanogap photonic platforms such as plasmonic cavities, metasurfaces, and hybrid microresonators to enable drift-tolerant population thermometry, in situ monitoring of local dielectric/gap variations under strain or temperature cycling, and calibrated readout of weak radiators beyond excitons (e.g., defect emitters or molecular transitions) in complex environments.

**Funding.** National Natural Science Foundation of China (12074259, 62122054).

**Disclosures.** The authors declare that there are no conflicts of interest related to this article.

**Data availability.** Data underlying the results presented in this paper are not publicly available at this time but may be obtained from the authors upon reasonable request.

**Supplemental document.** See Supplement 1 for supporting content.